\documentclass[twocolumn]{aastex631}
\usepackage{amsmath,amssymb}
\usepackage{chemformula}

\begin{document}

\title{On the abiotic origin of dimethyl sulfide: discovery of DMS in the Interstellar Medium}

\author[0000-0001-9629-0257]{Miguel Sanz-Novo}
\author[0000-0002-2887-5859]{V\'ictor M. Rivilla}
\affiliation{Centro de Astrobiolog{\'i}a (CAB), INTA-CSIC, Carretera de Ajalvir km 4, Torrej{\'o}n de Ardoz, 28850 Madrid, Spain}

\author[0000-0002-7104-8795]{Christian P. Endres}
\author[0000-0002-9890-5451]{Valerio Lattanzi}
\affiliation{Center for Astrochemical Studies, Max-Planck-Institut f\"{u}r Extraterrestrische Physik, Gießenbachstraße~1, 85748 Garching, Germany}

\author[0000-0003-4493-8714]{Izaskun Jim\'enez-Serra}
\affiliation{Centro de Astrobiolog{\'i}a (CAB), INTA-CSIC, Carretera de Ajalvir km 4, Torrej{\'o}n de Ardoz, 28850 Madrid, Spain}

\author[0000-0001-8064-6394]{Laura Colzi}
\affiliation{Centro de Astrobiolog{\'i}a (CAB), INTA-CSIC, Carretera de Ajalvir km 4, Torrej{\'o}n de Ardoz, 28850 Madrid, Spain}

\author[0000-0003-3721-374X]{Shaoshan Zeng}
\affiliation{Star and Planet Formation Laboratory, Cluster for Pioneering Research, RIKEN, 2-1 Hirosawa, Wako, Saitama, 351-0198, Japan}

\author[0000-0002-6389-7172]{Andr\'es Meg\'ias}
\affiliation{Centro de Astrobiolog{\'i}a (CAB), INTA-CSIC, Carretera de Ajalvir km 4, Torrej{\'o}n de Ardoz, 28850 Madrid, Spain}

\author[0000-0001-6049-9366]{\'Alvaro L\'opez-Gallifa}
\affiliation{Centro de Astrobiolog{\'i}a (CAB), INTA-CSIC, Carretera de Ajalvir km 4, Torrej{\'o}n de Ardoz, 28850 Madrid, Spain}

\author[0000-0001-5191-2075]{Antonio Mart\'inez-Henares}
\affiliation{Centro de Astrobiolog{\'i}a (CAB), INTA-CSIC, Carretera de Ajalvir km 4, Torrej{\'o}n de Ardoz, 28850 Madrid, Spain}

\author[0000-0001-7535-4397]{David San Andr\'es}
\affiliation{Centro de Astrobiolog{\'i}a (CAB), INTA-CSIC, Carretera de Ajalvir km 4, Torrej{\'o}n de Ardoz, 28850 Madrid, Spain}

\author[0000-0002-4782-5259]{Bel\'en Tercero}
\affiliation{Observatorio Astron\'omico Nacional (OAN-IGN), Calle Alfonso XII, 3, 28014 Madrid, Spain}

\author[0000-0002-5902-5005]{Pablo de Vicente}
\affiliation{Observatorio de Yebes (OY-IGN), Cerro de la Palera SN, Yebes, Guadalajara, Spain}

\author[0000-0001-9281-2919]{Sergio Mart\'in}
\affiliation{European Southern Observatory, Alonso de C\'ordova 3107, Vitacura 763 0355, Santiago, Chile}
\affiliation{Joint ALMA Observatory, Alonso de C\'ordova 3107, Vitacura 763 0355, Santiago, Chile}

\author[0009-0009-5346-7329]{Miguel A. Requena-Torres}
\affiliation{Department of Physics, Astronomy and Geosciences, Towson University, Towson, MD 21252, USA}

\author[0000-0003-1481-7911]{Paola Caselli}
\affiliation{Center for Astrochemical Studies, Max-Planck-Institut f\"{u}r Extraterrestrische Physik, Gießenbachstraße~1, 85748 Garching, Germany}

\author[0000-0003-4561-3508]{Jes\'us Mart\'in-Pintado}
\affiliation{Centro de Astrobiolog{\'i}a (CAB), INTA-CSIC, Carretera de Ajalvir km 4, Torrej{\'o}n de Ardoz, 28850 Madrid, Spain}

\begin{abstract}

Following the discovery of dimethyl sulfide (\ch{CH3SCH3}, DMS) signatures in comet 67P/Churyumov-Gerasimenko, we report the first detection of this organosulfur species in the interstellar medium, during the exploration of an ultradeep molecular line survey performed toward the Galactic Center molecular cloud G+0.693-0.027 with the Yebes 40$\,$m and IRAM 30$\,$m telescopes. We derive a molecular column density of $N$ = (2.6 $\pm$ 0.3)$\times$10$^{13}$ cm$^{-2}$, yielding a fractional abundance relative to H$_2$ of $\sim$1.9$\times$10$^{-10}$. This implies that DMS is a factor of $\sim$1.6 times less abundant than its structural isomer \ch{CH3CH2SH} and $\sim$30 times less abundant than its O-analogue dimethyl ether (\ch{CH3OCH3}) toward this cloud, in excellent agreement with previous results on various O/S pairs. Furthermore, we find a remarkable resemblance between the relative abundance of DMS/\ch{CH3OH} in G+0.693-0.027 ($\sim$1.7$\times$10$^{-3}$) and in the comet ($\sim$1.3$\times$10$^{-3}$). Although the chemistry of DMS beyond Earth is yet to be fully disclosed, this discovery provides conclusive observational evidence on its efficient abiotic production in the interstellar medium, casting doubts about using DMS as a reliable biomarker in exoplanet science.

\end{abstract}

\section{Introduction} 
\label{sec:intro}

In recent years, significant efforts are directed toward the search for signs of life beyond our Solar System via remote sensing of atmospheric biosignatures in habitable exoplanets. These biosignatures, also known as biomarkers, are seen as unambiguous molecular indicators of carbon-based life on Earth \citep{Sagan1993,Seager2016}. They include molecules produced by biological activity, ranging from the very simple O$_2$, a product of photosynthesis, to relatively more complex sulfur-bearing species such as dimethyl sulfide (DMS; \ch{CH3SCH3}), also known as dimethyl thioether or methylthiomethane. Atmospheric DMS on Earth is thought to originate solely from marine biological activity \citep{Pilcher2004}, through the degradation of dimethylsulfoniopropionate (DMSP; \ch{C5H10O2S}), a chemical compound mainly produced by marine phytoplankton, with no known natural abiotic sources identified so far on our planet or in space \citep{Hanni2024}. Consequently, DMS has been proposed as a potentially observable indicator of aquatic life on habitable exoplanets with a H$_2$-rich atmospheres \citep{Seager2013,Catling2018}, including the so-called “Hycean” worlds that could have habitable oceans \citep{Madhusudhan2021}. Among the best-known Hycean world candidate is the exoplanet K2-18b \citep{Benneke2019}, which was recently observed with the James Webb Space Telescope (JWST) NIRISS and NIRSpec instruments \citep{Madhusudhan2023b}, although the occurrence of Hycean conditions in K2-18b are currently under debate \citep{Wogan2024,Glein2024}. These JWST observations featured strong evidence of \ch{CH4} and \ch{CO2}, no signs of \ch{NH3}, and a tentative assignment of DMS, prompting the scientific community to investigate the chemistry of DMS beyond Earth.

The interstellar medium (ISM) is known to have an active sulfur (S) chemistry and therefore, the detection of DMS in the ISM would represent a clear abiotic source of this compound in exoplanets. Indeed, the comparison between the molecular abundances of diverse molecular species measured in star-forming regions and in comets are rather consistent, which suggests that minor bodies of the Solar System (i.e., asteroids, comets, and meteorites) may inherit the chemical composition of the parental interstellar cloud (i.e., pre-Solar environments; \citealt{Altwegg2015,altwegg2016,Altwegg2017,Drozdovskaya2018,rivilla2020,rivilla2021a,Hanni2022,Zeichner2023,Gallifa2024}). Very recently, the analysis of the data from the $\textit{Rosetta}$ Orbiter Instrument for Ion and Neutral Analysis (ROSINA) has provided strong evidence of the presence of DMS in the pristine cometary material of comet 67P/Churyumov-Gerasimenko (hereafter 67P; \citealt{Hanni2024}). This provides the first proof for the existence of an abiotic synthetic pathway to DMS in cometary matter, and highlights the urge to search for it in the ISM.

Recent observational findings in the Galactic Center (GC) molecular cloud G+0.693-0.027 (hereafter G+0.693) suggest that, unlike in star-forming regions and many other molecular clouds -where S tends to remain locked on the surface of dust grains (see e.g., \citealt{JimenezEscobar:2014kt,martin-domenech16,Laas2019,Fuente23})-, S appears to be less depleted \citep{Sanz-Novo2024a}. This mitigated depletion is likely due to increased sputtering erosion of dust particles driven by large-scale low-velocity shocks \citep{requena-torres_organic_2006,zeng2020} and implies that G+0.693 is an ideal laboratory for searching for DMS and studying S-chemistry. Indeed, among the more than 20 interstellar molecules detected toward this source in the last few years (see e.g., \citealt{rivilla2022a,jimenez-serra2022,SanzNovo23,Rivilla23,SanAndres2024}), several S-bearing species have been found, such as monothioformic acid (HC(O)SH), ethyl mercaptan (also known as ethanethiol, \ch{CH3CH2SH}; \citealt{rodriguez-almeida2021a}), O-protonated cabonyl sulfide (HOCS$^+$; \citealt{Sanz-Novo2024a}), thionylimide (HNSO; \citealt{Sanz-Novo2024b}) and the metal-bearing molecules MgS and NaS \citep{Rey-Montejo2024}.

In this letter, we report the discovery of DMS in the ISM toward G+0.693. Although this source does not show any signposts of active star formation (e.g., ultracompact H \textsc{ii} regions or dust continuum point sources \citealt{Ginsburg2018}), it is thought to host a shock-triggered prestellar condensation that might be on the verge of gravitational collapse and could eventually host protostars \citep{Colzi2024}. This fact is also supported by observations of deuterium fractionation (D/H ratios) toward G+0.693, which revealed the presence of a colder, denser, and less turbulent narrow gas component embedded in the warm and highly turbulent cloud \citep{Colzi2022}. Therefore, our study shows that DMS can be abiotically and efficiently formed in space, even in the early stages of star formation, and we stress the need for caution in using DMS as a unique biomarker in exoplanet science.

\begin{center}
\begin{figure*}[ht]
     \centerline{\resizebox{0.8
     \hsize}{!}{\includegraphics[angle=0]{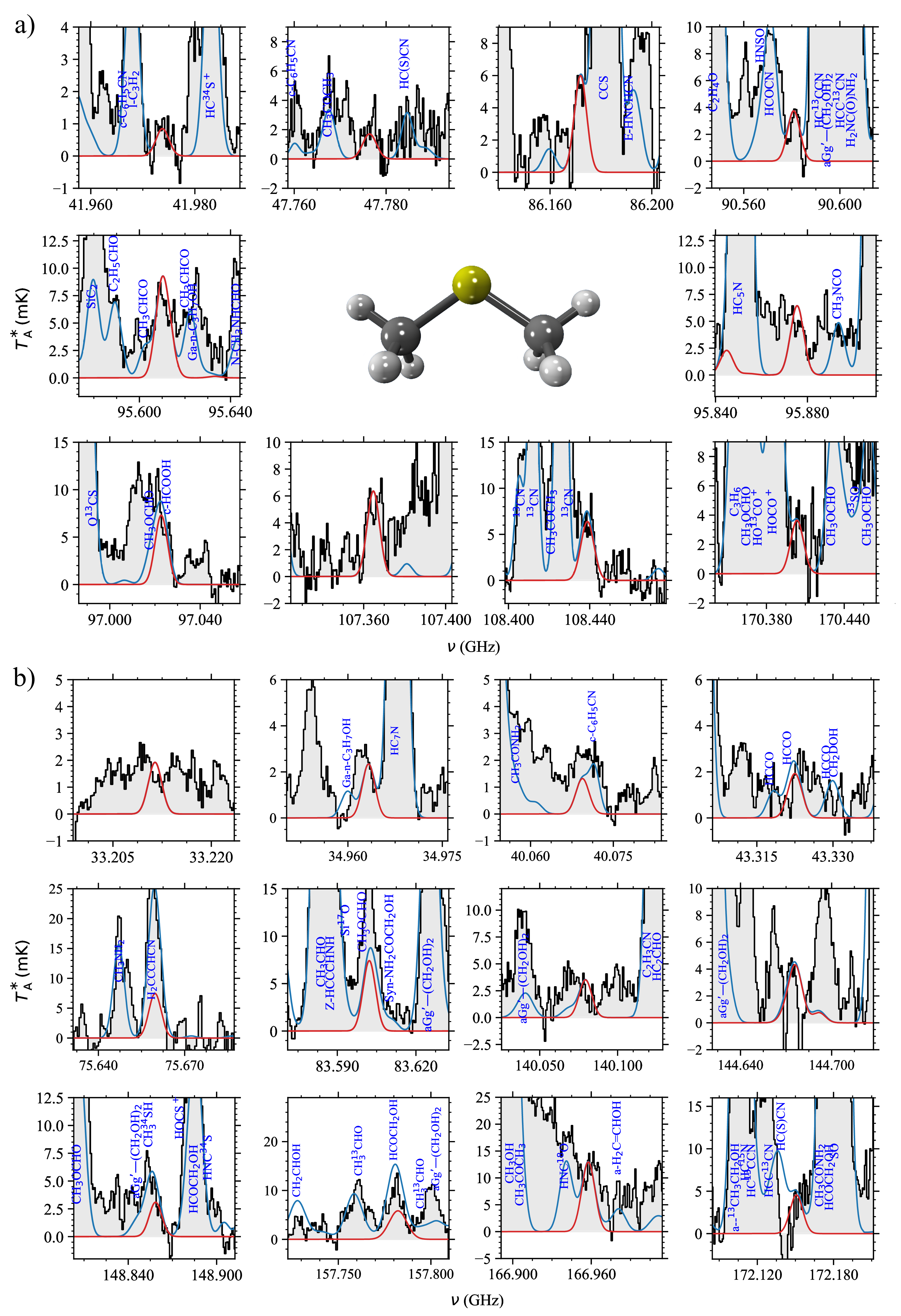}}}
     \caption{(a) Selected transitions of DMS identified toward the GC molecular cloud G+0.693–0.027, which were used to derive the LTE physical parameters of the molecule (see text; listed in Table \ref{tab:DMS}). (b) Additional transitions of DMS observed in the astronomical dataset that further support the detection but were excluded from the fit because of a non-negligible line blending with either known species or U-lines (listed in Table \ref{tab:DMSbis}; see Figures \ref{f:LTEall1} and \ref{f:LTEall2} for the rest of observed lines). The red line shows the best LTE fit of DMS, while the blue line represents the combined emission of all molecules identified in the survey, including DMS (observed spectra are shown as black histograms with a gray shaded area). The transitions are sorted by increasing frequency. The structure of DMS, optimized at the B2PLYPD3/aug-cc-pVTZ level using Gaussian 16 \citep{Frisch2016}, is also shown (carbon atoms in gray, sulfur atom in yellow and hydrogen atoms in white).}
\label{f:LTEspectrum}
\end{figure*}
\end{center}

\section{Observations} 
\label{sec:obs}

We have employed an unbiased and ultradeep spectral line survey carried out toward the GC molecular cloud G+0.693 to search for DMS. In particular, we have used $Q$-band observations at centimeter wavelengths (31.075-50.424 GHz), conducted with the Yebes 40-m radiotelescope (Guadalajara, Spain), as well as additional observations across two distinct millimeter-wavelength frequency windows (83.2–115.41 GHz and 142.00–173.81 GHz), conducted with the 30-m  IRAM radio telescope (Granada, Spain). We have used the position switching mode, centered at $\alpha$ = $\,$17$^{\rm h}$47$^{\rm m}$22$^{\rm s}$, $\delta$ = $\,-$28$^{\circ}$21$^{\prime}$27$^{\prime\prime}$, with the off position shifted by $\Delta\alpha$~=~$-885$$^{\prime\prime}$ and $\Delta\delta$~=~$290$$^{\prime\prime}$. The half power beam width (HPBW) of the Yebes 40$\,$m telescope varies between $\sim$35$^{\prime\prime}$-55$^{\prime\prime}$ (at 50 and 31 GHz, respectively; \citealt{tercero2021}) and the HPBW of the IRAM 30$\,$m radiotelescope is $\sim$14$^{\prime\prime}$$-$29$^{\prime\prime}$ in the observed frequency range. Also, we assumed that the molecular emission toward G+0.693 is extended as compared to the telescope beam \citep{Jones2012,Li2020,Zheng2024} and, therefore, the spectra are presented in units of antenna temperature ($T_A^*$). Further details on these observations, including resolution and final noise levels of the molecular line survey, are given in \citet{Rivilla23} and \citet{SanzNovo23}.

\section{Detection of DMS} 
\label{sec:detection}

\begin{table*}
\tabcolsep 2pt
\centering
\caption{Spectroscopic information of the selected transitions of DMS detected toward G+0.693$-$0.027, shown in Figure \ref{f:LTEspectrum}(a).}
\begin{tabular}{cccccccccccccc}
\hline
Frequency & Transition$^{(a)}$ & log \textit{I} (300 K) & $E$$\mathrm{_{up}}$  & rms & $\int$ $T$$\mathrm{_A^*}$d$v$ & S/N $^{(b)}$ & Blending  \\ 
(GHz) &                        &  (nm$^2$ MHz)           &  (K)      &  (mK) & (mK km s$^{-1}$)  & &    \\
\hline
41.9724761 (3) & 9$_{3,6}$--9$_{2,7}$(EA) & --5.9773 & 34.2 & 0.6 & 20 & 6 & Unblended\\
41.9724785 (3) & 9$_{3,6}$--9$_{2,7}$(AE) & --6.1526 & 34.2 & & & & Unblended\\
41.9732147 (2) & 9$_{3,6}$--9$_{2,7}$(EE) & --5.5508 & 34.2 & & & & Unblended\\
41.9739521 (3) & 9$_{3,6}$--9$_{2,7}$(AA) & --5.7546 & 34.2 & & & & Unblended\\
47.7751079 (4) & 6$_{3,3}$--6$_{2,4}$(EA) & --6.6187 & 18.3 & 0.9 & 55 & 10  & Unblended\\
47.7751471 (4) & 6$_{3,3}$--6$_{2,4}$(AE) & --6.3177 & 18.3 & & & & Unblended\\
47.7760787 (4) & 6$_{3,3}$--6$_{2,4}$(EE) & --5.7143 & 18.3 & & & & Unblended\\
47.7770299 (5) & 6$_{3,3}$--6$_{2,4}$(AA) & --6.1413 & 18.3 & & & & Unblended\\
86.1712950 (5) & 7$_{1,7}$--6$_{0,6}$(AE) & --5.6232 & 16.8 & 1.2 & 67 & 10 & Unblended\\
86.1712952 (5) & 7$_{1,7}$--6$_{0,6}$(EA) & --5.9242 & 16.8 & & & & Unblended\\
86.1714051 (5) & 7$_{1,7}$--6$_{0,6}$(EE) & --5.0244 & 16.8 & & & & Unblended\\
86.1715151 (5) & 7$_{1,7}$--6$_{0,6}$(AA) & --5.4476 & 16.8 & & & & Unblended\\
90.5795461 (5) & 5$_{2,4}$--4$_{1,3}$(AE) & --5.9670 & 11.6 & 1.3 & 52 & 7 & Unblended \\
90.5795502 (5) & 5$_{2,4}$--4$_{1,3}$(EA) & --6.2659 & 11.6 & & & & Unblended\\
90.5801981 (5) & 5$_{2,4}$--4$_{1,3}$(EE) & --5.3650 & 11.6 & & & & Unblended\\
90.5808481 (5) & 5$_{2,4}$--4$_{1,3}$(AA) & --5.7902 & 11.6 & & & & Unblended\\
95.6075736 (12) & 3$_{3,1}$--2$_{2,0}$(AE) & --5.9443 & 8.6 & 1.0 & 216 & 37 & Unblended\\
95.6087444 (6) & 8$_{0,8}$--7$_{1,7}$(EA) & --5.7703 & 21.4 & & & & Unblended\\
95.6087445 (6) & 8$_{0,8}$--7$_{1,7}$(AE) & --5.4686 & 21.4 & & & & Unblended\\
95.6087928 (6) & 8$_{0,8}$--7$_{1,7}$(EE) & --4.8669 & 21.4 & & & & Unblended\\
95.6088410 (6) & 8$_{0,8}$--7$_{1,7}$(AA) & --5.2927 & 21.4 & & & & Unblended\\
95.6108320 (6) & 3$_{3,1}$--2$_{2,0}$(EA) & --6.1816 & 8.6 & & & & Unblended\\
95.6110973 (6) & 3$_{3,1}$--2$_{2,0}$(EE) & --5.2992 & 8.6 & & & & Unblended\\
95.6131645 (6) & 3$_{3,1}$--2$_{2,0}$(AA) & --5.7044 & 8.6 & & & & Unblended\\
95.8728914 (6) & 3$_{3,0}$--2$_{2,1}$(EA) & --5.7034 & 8.6 & 1.0 & 96 & 18 & Unblended \\
95.8749605 (6) & 3$_{3,0}$--2$_{2,1}$(EE) & --5.2979 & 8.6 & & & & Unblended \\
95.8752278 (6) & 3$_{3,0}$--2$_{2,1}$(AA) & --5.4810 & 8.6 & & & & Unblended \\
95.8761497 (8) & 3$_{3,0}$--2$_{2,1}$(AE) & --5.9441 & 8.6 & & & & Unblended \\
97.0220250 (6) & 8$_{1,8}$--7$_{0,7}$(AE) & --5.4544 & 21.4 & 1.0 & 111 & 20 & Slightly blended: $c$-HCOOH, CH$_3$OCHO\\
97.0220251 (6) & 8$_{1,8}$--7$_{0,7}$(EA) & --5.2785 & 21.4 & & & & Slightly blended: $c$-HCOOH, CH$_3$OCHO \\
97.0221172 (6) & 8$_{1,8}$--7$_{0,7}$(EE) & --4.8523 & 21.4 & & & & Slightly blended: $c$-HCOOH, CH$_3$OCHO \\
97.0222095 (6) & 8$_{1,8}$--7$_{0,7}$(AA) & --5.0557 & 21.4 & & & & Slightly blended: $c$-HCOOH, CH$_3$OCHO \\
107.3639215 (6) & 9$_{0,9}$--8$_{1,8}$(EA) & --5.1352 & 26.5 & 1.6 & 149 & 16 & Unblended\\
107.3639215 (6) & 9$_{0,9}$--8$_{1,8}$(AE) & --5.3123 & 26.5 & & & & Unblended\\
107.3639780 (6) & 9$_{0,9}$--8$_{1,8}$(EE) & --4.7102 & 26.5 & & & & Unblended\\
107.3640344 (6) & 9$_{0,9}$--8$_{1,8}$(AA) & --4.9150 & 26.5 & & & & Unblended\\
108.4365514 (7) & 4$_{3,2}$--3$_{2,1}$(AE) & --5.7745 & 11.2 & 1.6 & 176 & 19 & Unblended\\
108.4370865 (6) & 4$_{3,2}$--3$_{2,1}$(EA) & --5.5971 & 11.2 & & & & Unblended\\
108.4381296 (6) & 4$_{3,2}$--3$_{2,1}$(EE) & --5.1710 & 11.2 & & & & Unblended\\
108.4394410 (6) & 4$_{3,2}$--3$_{2,1}$(AA) & --5.3750 & 11.2 & & & & Unblended\\
170.4005084 (9) & 7$_{4,4}$--6$_{3,3}$(AE) & --5.2517 & 26.4 & 1.0 & 69 & 12 & Unblended\\ 
170.4010524 (9) & 7$_{4,4}$--6$_{3,3}$(EA) & --5.5513 & 26.4 & & & & Unblended\\   
170.4024919 (8) & 7$_{4,4}$--6$_{3,3}$(EE) & --4.6510 & 26.4 & & & & Unblended\\   
170.4042038 (9)&  7$_{4,4}$--6$_{3,3}$(AA) & --5.0751 & 26.4 & & & & Unblended\\   
\hline 
\end{tabular}
\label{tab:DMS}
\vspace*{-0.5ex}
\tablecomments{$^{(a)}$ The rotational energy levels are labeled using the conventional notation for asymmetric tops: $J_{K_{a},K_{c}}$. 
The AA, EE, EA and AE labels refer to the different symmetry substates, arising from the presence of two methyl internal rotation motions. $^{(b)}$ The S/N ratio is computed from the integrated signal ($\int$ $T$$\mathrm{_A^*}$d$v$) and noise level, $\sigma$ = rms $\times$ $\sqrt{\delta v \times \mathrm{FWHM}}$, where $\delta$$v$ is the velocity resolution of the spectra and the FWHM is fitted from the data. Numbers in parentheses represent the predicted uncertainty associated to the last digits. We denote as “unblended" lines those that are not contaminated by other species (see Sect. \ref{sec:detection} for a description of blending criteria), although all the transitions belonging to different symmetry substates are partially or fully coalesced (auto-blended), and we provide the integrated intensity and S/N ratio of the mean observed line.}
\end{table*}

The rotational spectrum of DMS, was first investigated by \citet{rudolph1960}. Since then, the microwave, millimeter- and submillimeter-wave spectroscopy of DMS has been thoroughly analyzed by numerous research groups (see e.g., \citealt{dreizler1962,demaison1980,Jabri2016,Ilyushin2020}). To facilitate the interstellar search for DMS, we have prepared a new catalogue using the most updated available high-resolution data (see further details in Appendix \ref{rot_backgr}), which has been subsequently implemented into the \textsc{Madcuba} software package \citep{martin2019}.

To analyze the astronomical data, we employed the Spectral Line Identification and Modeling (SLIM) tool (version from 2023, November 15) within \textsc{Madcuba}. This tool allows us to generate synthetic spectra of DMS under the Local Thermodynamic Equilibrium (LTE) assumption, which can be directly compared with the observed spectrum, but also to evaluate the emission of all the molecules previously identified in the current spectral survey (more than 130 species). 

As shown in Figure \ref{f:LTEspectrum}(a), we have detected a total of ten clean $b$-type lines with an integrated signal-to-noise (S/N) ratio $\geq$6, belonging to both $R$- and $Q$-branches, which are listed in Table \ref{tab:DMS}. These lines were selected following the same criteria described in \citet{Rey-Montejo2024} to define unblended lines, addressing the plausible contamination by a yet unidentified (U) line. In essence, we analyzed the vicinity of the DMS lines within a velocity range of $\pm$FWHM/2, where FWHM corresponds to the observed linewidth. To quantify the contribution of the contamination to the overall area, we subtracted the LTE fit of the DMS from the observed spectrum. If the residual area contributes $\leq$25$\%$ to the total, the line is considered unblended. For instance, for the line at $\sim$95.87 GHz, the residual area contributes in 10.8\% to the total area, which is well below the selected threshold. In addition, if a contamination by a known molecule lies within this range but contributes less than 25$\%$ to the overall integrated intensity, we will consider it as slightly blended. Therefore, the selected subset of lines depicted in Figure \ref{f:LTEspectrum}(a) are all unblended, as they meet the criteria outlined above, with only one slightly blended line. The detection is further supported by the observation of additional lines that are either well reproduced by the observations once the contribution with the emission from all the species previously detected toward G+0.693 is accounted for (e.g., first panels in the second and third row of Figure \ref{f:LTEspectrum}b), or that are contaminated with a U-line contributing $\geq$25$\%$ to the integrated area but DMS appears in most cases to be the dominant carrier of the spectral features (see the first two panels shown in Figure \ref{f:LTEspectrum}b). For the remaining lines, the predicted LTE emission for the lines that appear heavily blended with more intense transitions, or the derived S/N ratio is lower than 5, but is in all cases consistent with the observed spectra. In Figures \ref{f:LTEall1} and \ref{f:LTEall2} we report every single transition of DMS covered by the survey that appears above an integrated S/N$\sim$3, which corresponds to a total of $\sim$50 lines (i.e., about 190 transitions). We stress that no missing lines of DMS are observed within the whole survey. Also, most of the observed lines consist of fully coalesced quartets of transitions exhibiting the same quantum numbers but belonging to different symmetry substates (i.e., AA, EE, EA and AE), which cannot be resolved due to the typical broad linewidths of the molecular line emission measured toward this source (FWHM $\sim$ 15$-$20 km s$^{-1}$; \citealt{requena-torres_organic_2006,requena-torres_largest_2008,zeng2018,rivilla2022c}). Nevertheless, this “autoblending" greatly facilitates detection in this case, due to the enhanced intensity of the resulting line cluster.

We obtained the best LTE modeling for DMS through a nonlinear least-squares LTE fit of the clean subset of transitions given in Table \ref{tab:DMS} (shown in Figure \ref{f:LTEspectrum}a) and using the \textsc{Autofit} tool within SLIM \citep{martin2019}. Both the FWHM and radial velocity ($v$$_{\rm LSR}$) were fixed in the fit to 20 km s$^{-1}$ and 67 km s$^{-1}$, respectively, while the excitation temperature ($T_{\rm ex}$) and column density ($N$) were left as free parameters. We derived a molecular column density of $N$(DMS) = (2.6 $\pm$ 0.3) $\times$10$^{13}$ cm$^{-2}$. This value is translated into a fractional abundance with respect to molecular hydrogen of (1.9$\pm$ 0.4) $\times$ 10$^{-10}$, adopting a $N$(H$_{2}$) = 1.35$\times$10$^{23}$ cm$^{-2}$ from \citet{martin_tracing_2008}, derived by using C$^{18}$O as a total H$_2$ column density tracer and assuming a C$^{18}$O/H$_2$ abundance ratio of 1.7$\times$10$^{-7}$ \citep{frerking_relationship_1982}. Also, we derived a $T_{\rm ex}$ = 13 $\pm$ 3 K, in agreement with the sub-thermal excitation conditions found for many other molecules toward G+0.693. This sub-thermal excitation of the molecular emission stems from the relatively low H$_2$ volume densities of few times 10$^4$ cm$^{-3}$ measured in the cloud, obtaining very low excitation temperatures (i.e., $T_{\rm ex}$ = 5 - 20 K) despite the higher kinetic temperature of the gas ($T_{\rm kin}$ ranging between 70-140 K; see e.g. \citealt{requena-torres_organic_2006,zeng2018}). In contrast to hot cores sources, only low-energy rotational transitions belonging to the ground vibrational state of the molecules are observable in G+0.693, which significantly reduces the levels of line blending and line confusion. Although these low H$_2$ densities could yield non-LTE effects (e.g., weak maser amplification) in the molecular excitation of relatively large molecules such as DMS, these effects usually appear at very low frequencies (especially $<$30 GHz; \citealt{Faure:2014iu,Faure2018}). Furthermore, collisional rate coefficients for DMS are currently unavailable, leaving LTE analysis as the only viable method to determine its excitation conditions in G+0.693. Regarding the $v$$_{\rm LSR}$, we used a value that is consistent with that found for a myriad of S-bearing species toward this source (e.g., OCS, HNCS, HNSO, SO and NaS with $v$$_{\rm LSR}$ = 66.8 $\pm$ 0.1, 66.7 $\pm$ 3, 68 $\pm$ 2, 67.9 $\pm$ 0.3 and 66.5 $\pm$ 1.5 km s$^{-1}$, respectively; \citealt{Sanz-Novo2024a,Sanz-Novo2024b}). The fit of the individual transitions of DMS are shown in red lines in Figure \ref{f:LTEspectrum}, and the global fit considering all molecular species identified and analyzed toward G+0.693, including DMS, is depicted in blue lines, both overlaid with the observational data (in black histogram and gray-shaded areas).

\begin{figure}
\centerline{\resizebox{0.95\hsize}{!}{\includegraphics[angle=0]{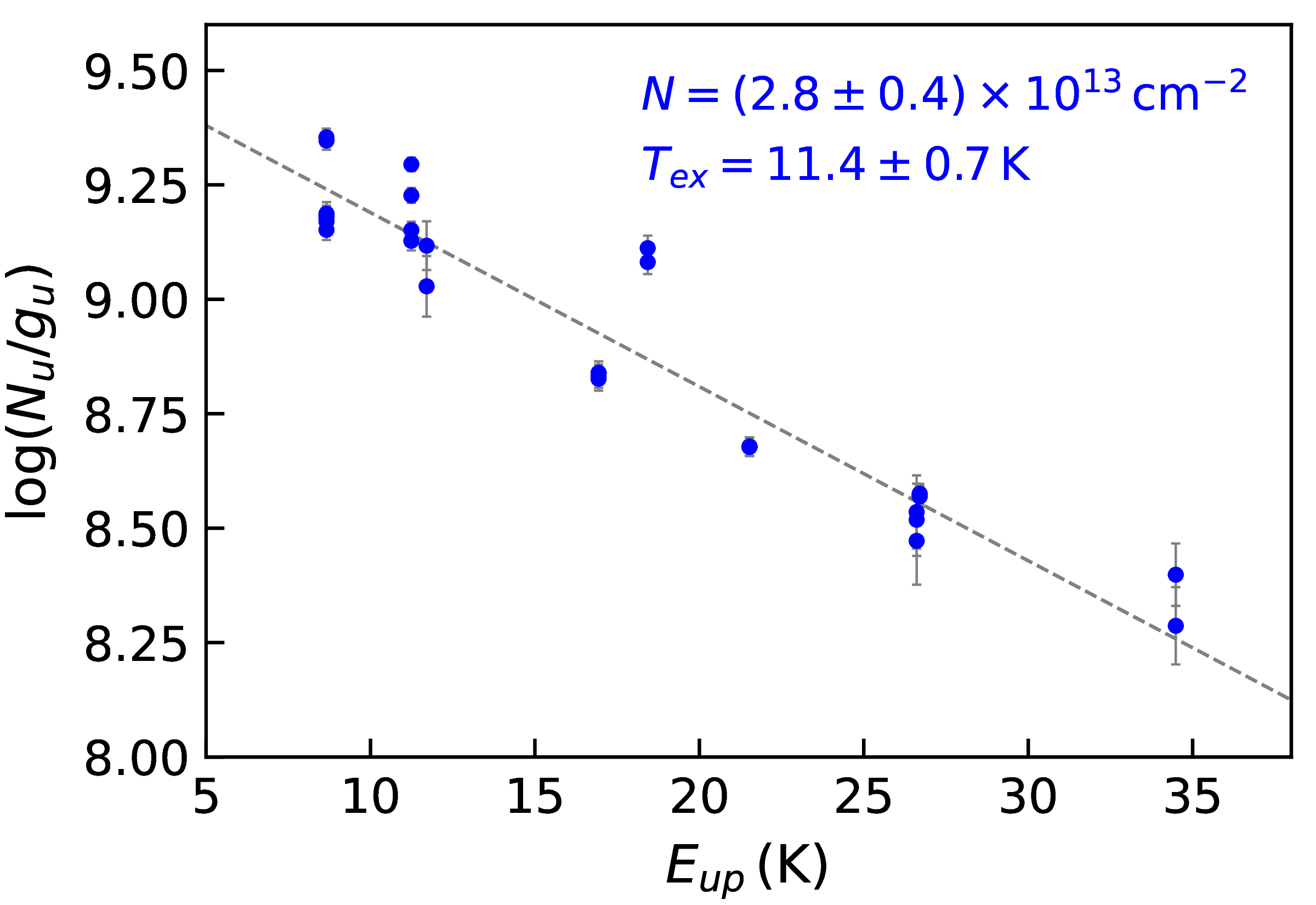}}}
\caption{Rotational diagram of DMS toward G+0.693 (blue dots, including 1$\sigma$ errors). The best linear fit to the data points is shown using a gray dotted line. The derived values for the molecular column density, $N$, and the excitation temperature, $T_{\rm ex}$, are shown in blue.} 
\label{f:rotdiagram}
\end{figure}

We also performed a complementary rotational diagram analysis \citep{goldsmith1999} using \textsc{Madcuba}. We employed all the transitions shown in Figure \ref{f:LTEspectrum}(a), whose spectroscopic information is listed in Table \ref{tab:DMS}), except for the slightly blended line, and considered the velocity-integrated intensity over the line width as described by \citet{rivilla2021a}. The results of the analysis are reported in Figure \ref{f:rotdiagram}. We obtained a $N$(DMS) = (2.8 $\pm$ 0.4) $\times$10$^{13}$ cm$^{-2}$ and $T_{\rm ex}$ = 11.4 $\pm$ 0.7 K, which are in good agreement with the physical parameters derived from the \textsc{Autofit}.

\begin{table*}
\centering
\caption{Derived physical parameters for S- and O-bearing molecules related to DMS detected toward G+0.693-0.027.}
\begin{tabular}{ c c c c c c c  c}
\hline
\hline
 Molecule & Formula & $N$   &  $T_{\rm ex}$ & $v$$_{\rm LSR}$ & FWHM  & Abundance$^a$ & Ref.$^b$   \\
 & & ($\times$10$^{13}$ cm$^{-2}$) & (K) & (km s$^{-1}$) & (km s$^{-1}$) & ($\times$10$^{-10}$)    \\
\hline
Dimethyl sufide (DMS) & CH$_3$SCH$_3$ & 2.6 $\pm$ 0.3 & 13 $\pm$ 3  &  67.0$^c$  & 20.0$^c$  & 1.9 $\pm$ 0.4 &  (1) \\
Ethyl mercaptan & $g$-\ch{CH3CH2SH} & 4 $\pm$ 2 & 10 $\pm$ 5  &  69.0$^c$  & 20.0$^c$  & 3 $\pm$ 1 &  (2) \\
Methyl mercaptan ($K$$_a$ = 0) & \ch{CH3SH} & 46.8 $\pm$ 0.5 & 8.5 $\pm$ 0.1  &  68.0 $\pm$ 0.1 & 21.2 $\pm$ 0.3  & 35 $\pm$ 5 &  (2) \\
Methanol  & \ch{CH3OH}  & - & - & - & - & 1100 $\pm$ 200$^d$ & (2) \\
Dimethyl ether (DME)  & \ch{CH3OCH3}  & 78 $\pm$ 3 & 18 $\pm$ 1 & 69.4 $\pm$ 0.5 & 20.6 $\pm$ 0.4 & 57 $\pm$ 9 & (1) \\
\hline
\end{tabular}
\label{tab:comparison}
\vspace{0mm}
\vspace*{-0.5ex}
\tablecomments{$^a$ We adopted $N_{\rm H_2}$ = 1.35$\times$10$^{23}$ cm$^{-2}$, from \citet{martin_tracing_2008}, assuming an uncertainty of 15\% of its value. $^b$ References: (1) This work; (2) \citet{rodriguez-almeida2021a}. $^c$ Values fixed in the fit. $^d$ Computed using the optically thin isotopologue CH$_3$$^{18}$OH and assuming $^{16}$O/$^{18}$O = 250 in the Central Molecular Zone (CMZ; \citealt{wilson_abundances_1994}).}
\label{tab:g0693}
\end{table*}

\section{Discussion} 
\label{sec:disc}

\begin{table*}
\centering
\tabcolsep 3pt
\caption{Relative abundances of DMS and related species in G+0.693-0.027 and comparison with comet 67P.}
\begin{tabular}{ c c c c c c }
\hline
\hline
 Source & DMS/\ch{CH3OH} ($\times$10$^{-3}$) & \ch{C2H6S}/\ch{CH3OH} ($\times$10$^{-3}$) & $g$-\ch{CH3CH2SH}/DMS & \ch{CH3SH}/\ch{CH3OH} ($\times$10$^{-2}$) & References$^c$ \\
\hline
G+0.693-0.027 & 1.7 $\pm$ 0.4 & 4.5 $\pm$ 2.0 $^a$ & 1.6 $\pm$ 0.6  & 3 $\pm$ 1 &   (1,2)\\
Comet 67P     &  -$^b$  & 1.3 $\pm$ 0.4 & - & 9 $\pm$ 3 &  (3)\\
\hline
\end{tabular}
\label{tab:abundance}
\vspace{0mm}
\vspace*{-0.5ex}
\tablecomments{$^a$ Computed taking the abundance of $g$-\ch{CH3CH2SH} into account, which was previously detected in  G+0.693 \citep{rodriguez-almeida2021a}. $^b$ Despite DMS is suggested as the most likely carrier for the $m$/$z$=62 ROSINA signal (see text; \citealt{Hanni2024}), these mass spectrometric measurements alone are in principle not able to differentiate between DMS and its isomers. $^c$ References: (1) This work; (2) \citet{rodriguez-almeida2021a}; (3) \citet{Hanni2024}.}
\end{table*}

The detection of DMS in the ISM confirms that an efficient abiotic formation pathway does occur under interstellar conditions. This finding underscores the need for caution when using DMS as a reliable biomarker in exoplanet science, providing an alternative explanation for its presence in the atmosphere of K2-12b \citep{Madhusudhan2023b}. However, to our knowledge, the formation pathways of DMS in the ISM are still poorly understood. To shed some light on its chemistry, we can initially contrast the molecular abundance of DMS reported in this work with the results obtained for structurally similar species. This comprises its O-bearing analogue, dimethyl ether (DME, \ch{CH3OCH3}), which is a well-known interstellar molecule \citep{Snyder:1974op,Lovas:1979in,requena-torres_organic_2006}, methanol (\ch{CH3OH}), along with other related S-bearing species also detected toward G+0.693, including its structural isomer \ch{CH3CH2SH} and the more simple \ch{CH3SH} \citep{rodriguez-almeida2021a}.

Based on the derived abundances and the ratios presented in Tables \ref{tab:comparison} and \ref{tab:abundance}, we find that DMS is a factor of 1.6 $\pm$ 0.6 less abundant than \ch{CH3CH2SH} toward this source. Moreover, we obtain a DME/DMS ratio of 30 $\pm$ 4. We can thus put these results in the context of a recent comprehensive study of the O/S ratio for a variety of well-known interstellar molecules across different interstellar environments (\citealt{Sanz-Novo2024a, Rey-Montejo2024} and ref. therein). The analysis presented in \cite{Sanz-Novo2024a} included cold molecular clouds, shock-dominated regions, high-mass and low-mass star-forming regions, as well as the comet 67P, aiming to stablish general trends across different stages of star formation. Overall, we found the lowest O/S ratio in G+0.693, where large-scale shocks are likely responsible for an increased sputtering erosion of dust particles \citep{requena-torres_organic_2006,zeng2020}, and in comet 67P, which are thought to exhibit relatively low S-depletion factors \citep{Calmonte16,Gallifa2024,Sanz-Novo2024a}. 

Interestingly, the DME/DMS abundance ratio in G+0.693 ($\sim$30) correlates very well with the general O/S trend observed in this source for related species (e.g., \ch{CH3OH}/\ch{CH3SH} and \ch{CH3CH2OH}/\ch{CH3CH2SH} abundance ratios of $\sim$31 and $\sim$15, respectively) and is also very close to the O/S solar value ($\sim$37; \citealt{Asplund:2009eu}). This behavior hints to the formation pathways of DMS resembling those of its O-analogue. Whereas \ch{CH3OH} and \ch{CH3CH2SH} most likely have a grain-surface origin \citep{watanabe2002,Qasim2018, Santos2024}, DME can be formed either on grain surfaces through the radical recombination reaction \citep{Garrod:2006gw}: \ch{CH3}  + \ch{CH3O} $\rightarrow$ \ch{CH3OCH3}, or in the gas phase through the two-reaction sequence between \ch{CH3OH} and protonated methanol (\ch{CH3OH2+}), followed by a proton transfer reaction \citep{Garrod:2006gw,taquet_formation_2016,skouteris2019}. As a result, the abundance of DME is expected to change by three orders of magnitude within a short period of time after \ch{CH3OH} is released into gas phase \citep{Caselli1993, Charnley:1995od}. Thus, if we assume that a similar chemistry takes place for DMS, we can propose the S-analogous reactions to be the most likely formation pathways, i.e. the  grain surface reaction \ch{CH3}  + \ch{CH3S} $\rightarrow$ \ch{CH3SCH3} or, alternatively, the gas phase routes starting from \ch{CH3SH} and \ch{CH3SH2+}, or \ch{CH3OH} and \ch{CH3SH2+}. Nonetheless, future theoretical and laboratory effort is certainly needed to delve deeper into the interstellar chemistry of DMS and assess the feasibility and prevalence of the aforementioned routes.


In this context, we can examine the relative abundance of DMS/\ch{CH3OH} in G+0.693 (see Table \ref{tab:abundance}), obtaining a ratio of (1.7 $\pm$ 0.4) $\times$10$^{-3}$, which can be contrasted with the value found in comet 67P under two possible scenarios \citep{Hanni2024}. First, if we assume that the $m$/$z$=62 ROSINA signal (\ch{C2H6S}) mainly belongs to DMS, the derived ratio in G+0.693 appears to be in remarkable agreement with the \ch{C2H6S}/\ch{CH3OH} value found in comet 67P of (1.3 $\pm$ 0.4) $\times$10$^{-3}$ \citep{Hanni2024}. These results align well with the conclusions of \citet{Hanni2024}, who re-analyzed and completed the high-resolution mass spectrometric data from \citet{Calmonte16} and \citet{Mahjoub23}, favoring the assignment of DMS over its structural isomer \ch{CH3CH2SH} based on a more compatible electron ionization induced fragmentation pattern. Additionally, they observed that the \ch{C2H6S}/\ch{CH3OH} ratio remained stable across various observational data sets, which probed several spacecraft's relative positions, and suggest that both DMS and \ch{CH3OH} sublimate from the same layer within the cometary nucleus. Alternatively, if we consider that both isomers, DMS and \ch{CH3CH2SH}, contribute to the signal measured in the comet, the derived \ch{C2H6S}/\ch{CH3OH} ratio of 4.5 $\pm$ 2 observed in G+0.693 is still within a factor of $\sim$3 relative to that obtained in comet 67P. Although additional correlation studies are needed \citep{Rubin2023,Gallifa2024}, the excellent match between G+0.693 and the cometary ratio, along with the points discussed earlier, suggest that both DMS and \ch{CH3OH} may be chemically linked throughout different phases of star-formation, from molecular clouds (i.e., G+0.693) to the end products such as comets and meteorites \citep{Hanni2024}.

In summary, the results presented in this letter provide compelling observational evidence that: i) DMS can be abiotically and efficiently formed in space, even in the early stages of star formation, which further strengthens the connection between the chemical inventory of the ISM and the rich feedstock of the minor bodies in the Solar System; ii) Although the interstellar chemical network involving DMS has not yet been fully unveiled, we emphasize the need for caution in using DMS as a reliable biomarker in exoplanet science; iii) The G+0.693 molecular cloud stands as an astrochemical treasure trove for the detection of new S-bearing species of increasing complexity levels, which are now accessible thanks to the superb sensitivity of current ultradeep molecular line surveys.

\software{1) Madrid Data Cube Analysis (\textsc{Madcuba}) on ImageJ is a software developed at the Center of Astrobiology (CAB) in Madrid; \url{https://cab.inta-csic.es/madcuba/}; \citet{martin2019}; version from 2023 November 15. 2) The 3D representation of DMS has been visualized with IQmol \url{https://iqmol.org/index.html}.}

\begin{acknowledgments}
 
We are grateful to the IRAM 30$\,$m and Yebes 40$\,$m telescopes staff for their help during the different observing runs, highlighting project 21A014 (PI: Rivilla), project 018-19 (PI: Rivilla) and projects 123-22 and 076-23 (PI: Jim\'enez-Serra). The 40$\,$m radio telescope at Yebes Observatory is operated by the Spanish Geographic Institute (IGN, Ministerio de Transportes, Movilidad y Agenda Urbana). IRAM is supported by INSU/CNRS (France), MPG (Germany) and IGN (Spain). M. S.-N. acknowledges a Juan de la Cierva Postdoctoral Fellow proyect JDC2022-048934-I, funded by the Spanish Ministry of Science, Innovation and Universities/State Agency of Research MICIU/AEI/10.13039/501100011033 and by the European Union “NextGenerationEU”/PRTR”. V. M. R.  acknowledges support from the grant RYC2020-029387-I funded by MICIU/AEI/10.13039/501100011033 and by “ESF, Investing in your future", from the Consejo Superior de Investigaciones Cient{\'i}ficas (CSIC) and the Centro de Astrobiolog{\'i}a (CAB) through the project 20225AT015 (Proyectos intramurales especiales del CSIC), and from the grant CNS2023-144464 funded by MICIU/AEI/10.13039/501100011033 and by “European Union NextGenerationEU/PRTR”. I. J.-S., J. M.-P., V. M. R., M .S.-N., L. C, A. M., A. L.-G and A. M. H. acknowledge funding from grant No. PID2022-136814NB-I00 from MICIU/AEI/10.13039/501100011033 and by “ERDF, UE A way of making Europe” .I.J.-S. also acknowledges funding from the ERC grant OPENS (project number 101125858) funded by the European Union. Views and opinions expressed are however those of the author(s) only and do not necessarily reflect those of the European Union or the European Research Council Executive Agency. Neither the European Union nor the granting authority can be held responsible for them. D.S.A. expresses his gratitude for the funds from the Comunidad de Madrid through Grant PIPF-2022/TEC-25475, and the financial support by the Consejo Superior de Investigaciones Cient{\'i}ficas (CSIC) and the Centro de Astrobiolog{\'i}a (CAB) through project 20225AT015 (Proyectos intramurales especiales del CSIC). B. T. thanks the Spanish MICIU for funding support from grants PID2022-137980NB-I00 and PID2023-147545NB-I00. S. Z. acknowledge the support by RIKEN Special Postdoctoral Researchers Program. C. P. E., V. L., and P.C. acknowledge the Max Planck society for the financial support.

\end{acknowledgments}

\bibliography{rivilla,bibliography}{}
\bibliographystyle{aasjournal}

\newpage
\appendix

\restartappendixnumbering
 
\section{Spectroscopic line catalogue of DMS}
\label{rot_backgr}

DMS appears as a challenge for rotational spectroscopy due to the presence of two equivalent methyl internal rotors that cause a splitting of each rotational level into the substates AA, AE, EA, EE of about a few Megahertz. Generally, the splitting between AE and EA cannot be resolved and a triplet of transitions is observed in the laboratory spectra. Its rotational spectrum was initially investigated by \citet{rudolph1960}. Afterward, the micro-, millimeter- and submillimeter-wave spectroscopy of DMS has been thoroughly analyzed by numerous research groups (see e.g., \citealt{dreizler1962,demaison1980,Jabri2016,Ilyushin2020}). To prepare the corresponding catalogue needed to guide the astronomical search for DMS, we used the most updated and complete dataset presented in \citet{Ilyushin2020}, converting Table S3 of that work into the common SPFIT/SPCAT catalogue format \citep{Pickett1991}. The intensity $I$(T) in $nm^2 MHz$ has been obtained using the reported value of $g_I S_g \mu_g^2$ (i.e., dipole moment squared multiplied by the transition linestrength) via:

{\small
\[
I(T) = \frac{8 \pi^3}{3hc} \nu g_I S_g \mu_g^2 \frac{\exp(-E''/kT) - \exp(-E'/kT)}{Q_{\text{r}}(T)}
\]
}

or rather

{\small
\[
I(T) = 4.16231 \cdot 10^{-5} \cdot \nu \cdot g_I \cdot S_g \cdot \mu_g^2 \cdot \frac{\exp(-E''/kT) - \exp(-E'/kT)}{Q_{\text{r}}(T)}
\]
}

adopting a dipole moment of 1.5 D, derived in previous Stark modulated measurements \citep{Pierce:1961bm}.

We have also computed the rotational partition function ($Q$$_r$) by direct summation of the ground state energy levels. We provide these values at the usual temperatures as implemented in the Jet Propulsion Lab (JPL) and the Cologne Database for Molecular Spectroscopy (CDMS) databases \citep{1998JQSRT..60..883P,Muller2005,endres2016}, including two additional temperatures of 2.725 K and 5.000 K (see Table \ref{t:pfun}).

\begin{table}[h!]
\begin{center}
\caption{Rotational partition function (Q$_r$) of DMS.}
\label{t:pfun}
\begin{tabular}{cc}
\hline
\multicolumn{1}{c}{Temperature}{\small (K)} & \multicolumn{1}{c}{Q$_r$} \\  
\hline
2.725 &	900.705 \\
5.000 &	2204.805 \\
9.375 &	5613.150 \\
18.75 &	15801.396 \\
37.50 &	44593.987 \\
75.00 &	126027.927 \\
150.0 &	356460.811 \\
225.0 &	653340.364 \\
300.0 &	995806.412 \\
\hline 
\end{tabular}
\end{center}
\vspace*{-2.5ex}
\end{table}

\restartappendixnumbering

\section{Complementary figures and tables}
\label{comp_fig}

\begin{center}
\begin{figure*}[ht]
     \centerline{\resizebox{1.0
     \hsize}{!}{\includegraphics[angle=0]{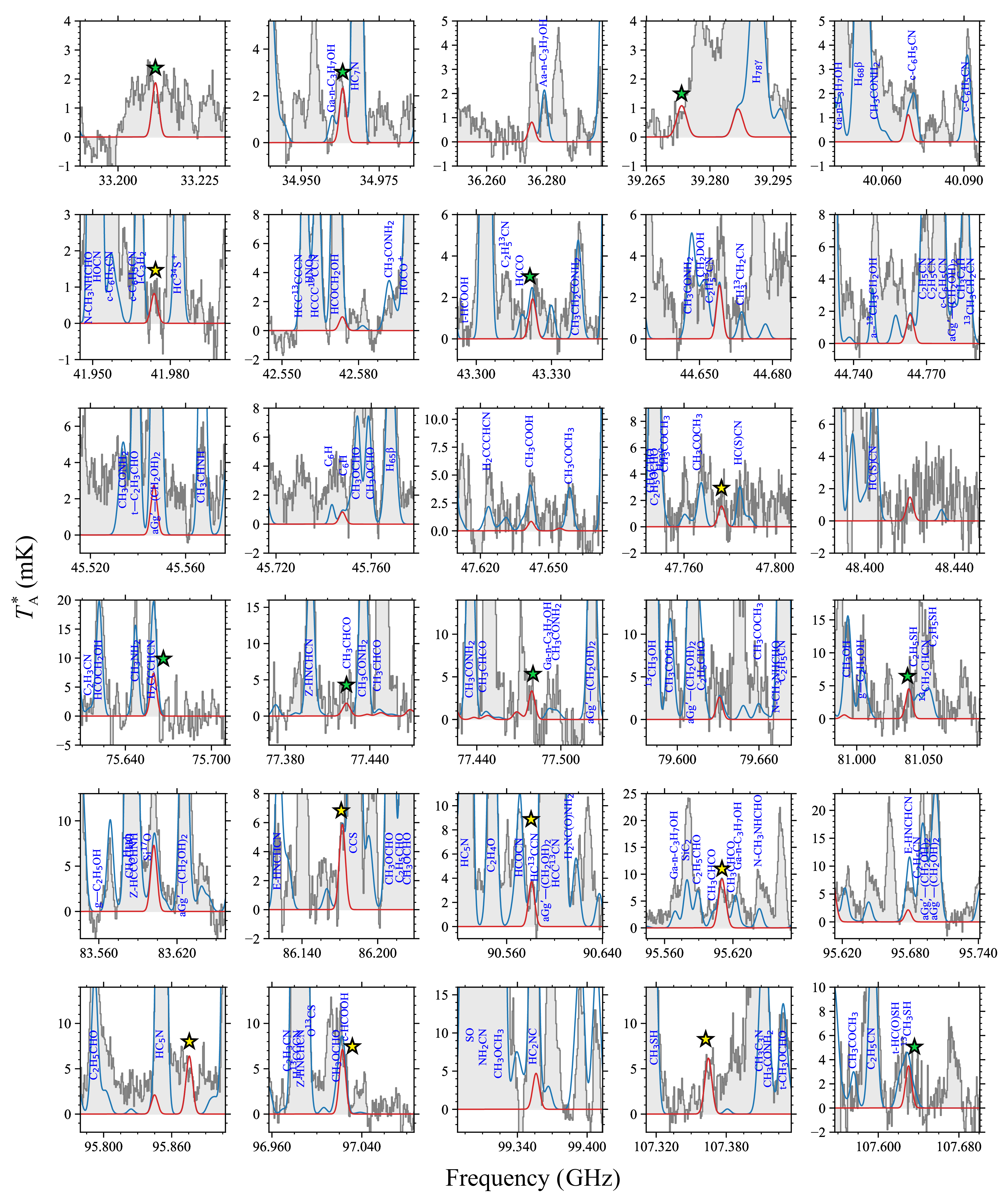}}}
     \caption{All the transitions of DMS observed toward the GC molecular cloud G+0.693–0.027 above an integrated S/N of $\sim$3, which include those used to derive the LTE physical parameters of the molecule highlighted with a yellow star (see main text; listed in Table \ref{tab:DMS}) together with many other transitions, marked with a green star, that are consistent with the observations but were excluded from the fit because of a non-negligible line blending with either known species or U-lines, or due to a lower S/N ratio. The result of the best LTE fit of DMS is plotted with a red line and the blue line depicts the emission from all the molecules identified to date in the survey, including DMS (observed spectra shown as gray histograms and light gray shaded area). The transitions are sorted by increasing frequency.}
\label{f:LTEall1}
\end{figure*}
\end{center}

\begin{center}
\begin{figure*}[ht]
     \centerline{\resizebox{1.0
     \hsize}{!}{\includegraphics[angle=0]{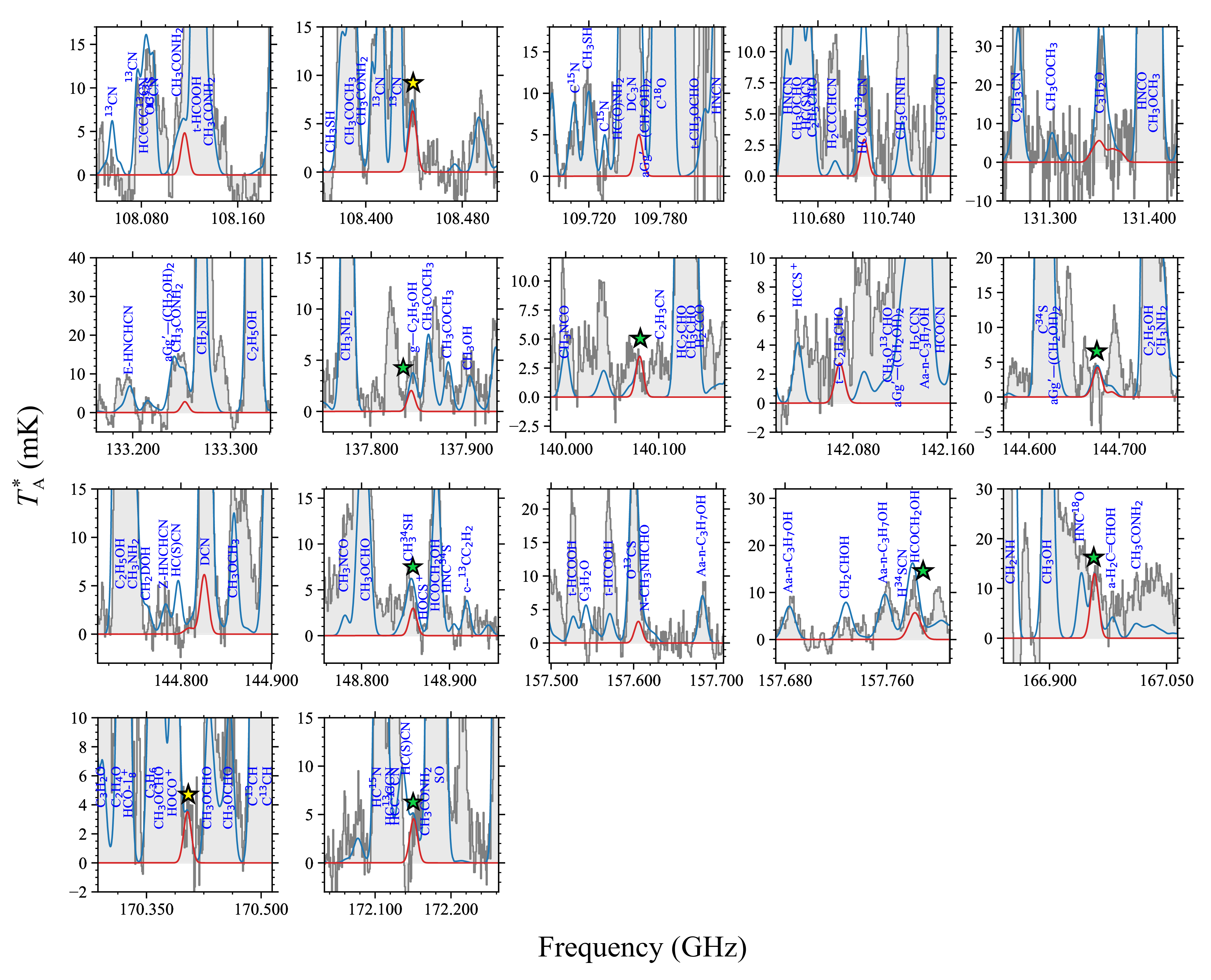}}}
     \caption{Same as in Figure \ref{f:LTEall1} but for additional lines.}
\label{f:LTEall2}
\end{figure*}
\end{center}

In Figures \ref{f:LTEall1} and \ref{f:LTEall2} we report every single transition of DMS covered by the survey conducted toward G+0.693 in three different frequency windows (i.e., between 31.075-50.424 GHz, 83.2-115.41 GHz and 142.0-173.81 GHz) that appear above an integrated of S/N$\sim$3. We also report those used to derive the physical properties of the molecule (marked with a yellow star) along with many other transitions (highlighted with a green star), that are consistent with the observations but were excluded from the fit because of a non-negligible line blending with either known species or U-lines, or due to a lower S/N ratio, but further strengthen the detection.

\clearpage

In Table \ref{tab:DMSbis} we provide the spectroscopic information for additional transitions of DMS also detected toward G+0.693$-$0.027 but that appear to be blended with the emission by other already identified molecules or contaminated by a yet unidentified features (shown in Figure \ref{f:LTEspectrum}b).

\begin{longtable*}{cccccc}
\caption{Additional transitions of DMS detected toward G+0.693$-$0.027, which are depicted in Figure \ref{f:LTEspectrum}(b).} \label{tab:DMSbis}\\
\hline 
Frequency & Transition$^{(a)}$ & log \textit{I} (300 K) & $E$$\mathrm{_{up}}$ & Blending \\ 
(GHz) &                        &  (nm$^2$ MHz)           &  (K)      &    \\
\hline
\endfirsthead
\caption[]{(Continued)} \\
\hline
Frequency & Transition$^{(a)}$ & log \textit{I} (300 K) & $E$$\mathrm{_{up}}$ & Blending \\ 
(GHz) &                        &  (nm$^2$ MHz)           &  (K)      &    \\
\hline
\endhead
\hline
\endfoot
33.2106615 (3) & 7$_{2,5}$--7$_{1,6}$(EA) & -6.2749 & 20.8 & U-line \\
33.2106618 (3) & 7$_{2,5}$--7$_{1,6}$(AE) & -6.4512 & 20.8 & U-line \\
33.2111965 (2) & 7$_{2,5}$--7$_{1,6}$(EE) & -5.8496 & 20.8 & U-line \\
33.2117342 (3) & 7$_{2,5}$--7$_{1,6}$(AA) & -6.0541 & 20.8 & U-line \\
34.9628152 (2) & 2$_{1,2}$--1$_{0,1}$(AE) & --6.9594 & 2.3 & U-line\\
34.9628167 (2) & 2$_{1,2}$--1$_{0,1}$(EA) & --6.7844 & 2.3 & U-line\\
34.9630536 (2) & 2$_{1,2}$--1$_{0,1}$(EE) & --6.3581 & 2.3 & U-line\\
34.9632912 (2) & 2$_{1,2}$--1$_{0,1}$(AA) & --6.5621 & 2.3 & U-line\\
40.0684823 (5) & 6$_{1,5}$--6$_{0,6}$(EA) & --6.9063 & 14.6 & $c$-\ch{C6H5CN}, U-line\\
40.0684827 (5) & 6$_{1,5}$--6$_{0,6}$(AE) & --6.6053 & 14.6 & $c$-\ch{C6H5CN}, U-line\\
40.0690965 (4) & 6$_{1,5}$--6$_{0,6}$(EE) & --6.0037 & 14.6 & $c$-\ch{C6H5CN}, U-line\\
40.0697104 (5) & 6$_{1,5}$--6$_{0,6}$(AA) & --6.4292 & 14.6 & $c$-\ch{C6H5CN}, U-line\\
43.3213865 (3) & 4$_{2,3}$--4$_{1,4}$(AA) & --6.7136 & 8.5 & HCCO, U-line\\
43.3213950 (3) & 4$_{2,3}$--4$_{1,4}$(EE) & --6.5375 & 8.5 & HCCO, U-line\\
43.3221966 (3) & 4$_{2,3}$--4$_{1,4}$(EA) & --6.1122 & 8.5 & HCCO, U-line \\
43.3230024 (3) & 4$_{2,3}$--4$_{1,4}$(AE) & -6.3156  & 8.5 & HCCO, U-line\\
75.6590519 (5) & 6$_{1,6}$--5$_{0,5}$(AE) & --5.8161 & 12.8 & \ch{H2CCCHCN}\\
75.6590522 (5) & 6$_{1,6}$--5$_{0,5}$(EA) & --5.6400 & 12.8 & \ch{H2CCCHCN}\\
75.6591878 (5) & 6$_{1,6}$--5$_{0,5}$(EE) & --5.2130 & 12.8 & \ch{H2CCCHCN}\\
75.6593216 (5) & 6$_{1,6}$--5$_{0,5}$(AA) & --5.4202 & 12.8 & \ch{H2CCCHCN}\\
83.6016039 (5) & 7$_{0,7}$--6$_{1,6}$(EA) & --5.4780 & 16.8 & \ch{CH3OCHO}, U-line \\
83.6016040 (5) & 7$_{0,7}$--6$_{1,6}$(AE) & --5.6538 & 16.8 & \ch{CH3OCHO}, U-line\\
83.6016375 (5) & 7$_{0,7}$--6$_{1,6}$(EE) & --5.0509 & 16.8 & \ch{CH3OCHO}, U-line\\
83.6016710 (5) & 7$_{0,7}$--6$_{1,6}$(AA) & --5.2550 & 16.8 & \ch{CH3OCHO}, U-line \\
140.0771605 (7) & 6$_{3,3}$--5$_{2,4}$(EA) & --5.8995 & 16.0 & Slightly blended: U-line\\    
140.0772051 (7) & 6$_{3,3}$--5$_{2,4}$(AE) & --5.5985 & 16.0 & Slightly blended: U-line\\ 
140.0784093 (7) & 6$_{3,3}$--5$_{2,4}$(EE) & --4.9959 & 16.0 & Slightly blended: U-line\\    
140.0796357 (7) & 6$_{3,3}$--5$_{2,4}$(AA) & --5.4224 & 16.0 & Slightly blended: U-line\\ 
144.6680673 (19) & 5$_{4,2}$--4$_{3,1}$(AE) & --5.5490 & 18.1 & Unblended\\
144.6742456 (9) & 5$_{4,2}$--4$_{3,1}$(EE) & --4.8541 & 18.1 & Unblended\\
144.6748530 (8) & 5$_{4,2}$--4$_{3,1}$(EA) & --5.6878 & 18.1 & Unblended\\
144.6780013 (8) & 5$_{4,2}$--4$_{3,1}$(AA) & --5.2097 & 18.1 & Unblended\\
148.8564866 (8) & 8$_{3,6}$--7$_{2,5}$(AE) & --5.4447 & 27.9 & CH$_3$$^{34}$SH \\
148.8564947 (8) & 8$_{3,6}$--7$_{2,5}$(EA) & --5.2692 & 27.9 & CH$_3$$^{34}$SH\\
148.8575630 (8) & 8$_{3,6}$--7$_{2,5}$(EE) & --4.8435 & 27.9 & CH$_3$$^{34}$SH\\
148.8586354 (9) & 8$_{3,6}$--7$_{2,5}$(AA) & --5.0478 & 27.9 & CH$_3$$^{34}$SH\\
157.7780248 (11) & 6$_{4,3}$--5$_{3,2}$(AE) & --5.3295 & 21.9 & \ch{HCOCH2OH} \\
157.7799851 (8) & 6$_{4,3}$--5$_{3,2}$(EA) & --5.1377 & 21.9 & \ch{HCOCH2OH}\\
157.7810616 (8) & 6$_{4,3}$--5$_{3,2}$(EE) & --4.7152 & 21.9 & \ch{HCOCH2OH}\\
157.7831399 (9) & 6$_{4,3}$--5$_{3,2}$(AA) & --4.9158 & 21.9 & \ch{HCOCH2OH}\\  
166.9526950 (10) & 5$_{5,1}$--4$_{4,1}$(AE) & --5.1561 & 22.8 & Slightly blended: U-line\\      
166.9539284 (9)  & 5$_{5,1}$--4$_{4,1}$(EA) & --5.4560 & 22.8 & Slightly blended: U-line\\     
166.9553062 (9)  & 5$_{5,1}$--4$_{4,1}$(EA) & --4.9789 & 22.8 & Slightly blended: U-line\\       
166.9555947 (8)  & 5$_{5,1}$--4$_{4,1}$(EE) & --4.5556 & 22.8 & Slightly blended: U-line\\    
166.9565396 (9)  & 5$_{5,1}$--4$_{4,1}$(AE) & --5.1561 & 22.8 & Slightly blended: U-line\\      
166.9574584 (8)  & 5$_{5,1}$--4$_{4,1}$(EE) & --4.5556 & 22.8 & Slightly blended: U-line\\  
166.9577469 (9)  & 5$_{5,1}$--4$_{4,1}$(AA) & --4.9789 & 22.8 & Slightly blended: U-line\\    
166.9591246 (9)  & 5$_{5,1}$--4$_{4,1}$(AA) & --4.7570 & 22.8 & Slightly blended: U-line\\ 
172.1479161 (9)  & 7$_{4,3}$--6$_{3,4}$(EA) & --5.0709 3 & 26.4 & HC(S)CN \\ 
172.1484601 (8)  & 7$_{4,3}$--6$_{3,4}$(AE) & --5.2484 3 & 26.4 & HC(S)CN \\  
172.1496314 (8)  & 7$_{4,3}$--6$_{3,4}$(EE) & --4.6456 3 & 26.4 & HC(S)CN \\  
172.1510742 (9)  & 7$_{4,3}$--6$_{3,4}$(AA) & --4.8485 3 & 26.4 & HC(S)CN \\
\end{longtable*}
\onecolumngrid
\tablecomments{$^{(a)}$ The rotational energy levels are labeled using the conventional notation for asymmetric tops: $J_{K_{a},K_{c}}$. Additionally, the AA,  EE, EA and AE labels refer to the different symmetry or torsional substates, arising from the presence of two methyl internal rotation motions. Numbers in parentheses represent the predicted uncertainty associated with the last digits. We denote as U-line the line blending with a yet unidentified feature.}

\end{document}